\documentclass[preprint,preprintnumbers,amsmath,amssymb]{revtex4}

\usepackage{graphicx}           
\usepackage{dcolumn}            
\usepackage{bm}                 
\usepackage{hyperref}

\begin{document}


\title{Efficient single-photon emission from electrically driven InP quantum dots epitaxially grown on Si(001)}

\author{M.~Wiesner$^{\star}$, W.-M.~Schulz, C.~Kessler, M.~Reischle, R.~Ro\ss bach,\\ M.~Jetter, \& P. Michler}

\affiliation{Institut f\"ur Halbleiteroptik und Funktionelle
Grenzfl\"achen and Research Center SCoPE, Universit\"at Stuttgart,
Allmandring 3, 70569
Stuttgart, Germany.\\ \\
$^{\star}$~Email correspondence: m.wiesner@ihfg.uni-stuttgart.de}

\maketitle

\textbf{The heteroepitaxy of III-V semiconductors on silicon is a
promising approach for making silicon a photonic platform for
on-chip optical interconnects\cite{ITRS, Mathine97, Miller09} and
quantum optical applications\cite{Politi08, O'Brien09, Almeida04}.
Monolithic integration of both material systems is a long-time
challenge, since different material properties lead to high defect
densities in the epitaxial layers. In recent years, nanostructures
however have shown to be suitable for successfully realising light
emitters on silicon\cite{Martensson04, Benyoucef09, Mi09,
Tanoto09, Chen11, Roest06}, taking advantage of their geometry.
Facet edges and sidewalls can minimise or eliminate the formation
of dislocations\cite{Mi09}, and due to the reduced contact area,
nanostructures are little affected by dislocation
networks\cite{Roest06}. Here we demonstrate the potential of
indium phosphide quantum dots as efficient light emitters on
CMOS-compatible silicon substrates, with luminescence
characteristics comparable to mature devices realised on III-V
substrates. For the first time, electrically driven single-photon
emission on silicon is presented, meeting the wavelength range of
silicon avalanche photo diodes' highest detection efficiency.}

To date, numerous realisations of light emitting diodes (LEDs) or
even lasers on silicon (Si) have been reported\cite{Liang10,
Groenert03, Kwon06}, mostly however using misoriented substrates
or thick buffer layers, which precludes a direct integration in
mainstream Si technology, as complementary metal oxide
semiconductor (CMOS) processes require exactly oriented Si(001)
substrates. Thick buffer layers cause large height differences and
thus negatively impact high resolution lithography. Alternatively,
III-V emitters have been realised on Si by hybrid
technologies\cite{Fang06, Roelkens10}, which however is a very
complex method, making CMOS integration time-consuming and
expensive.

The methods mentioned above are indispensable for providing
optimal growth conditions required for spatially extensive layers,
such as quantum wells. Defects within these layers deteriorate the
optical and electrical characteristics of the device and may lead
to reduced reliability. However, nanostructures were found to grow
defect-free even in a suboptimal environment\cite{Mi09} and thus
represent an enclosed electronic system. Using quantum dots (QDs)
as the active light emitting medium further provides the
possibility to build up semiconductor lasers with superior
performance, such as low thresholds and broader gain spectra. Over
and above, self-assembled (QDs) are excellent candidates for
single-photon sources, which are essential for quantum information
science and technology\cite{Michler00b, Yuan02, Santori02}.

Our self-assembled InP QDs were grown on exactly oriented Si(001)
substrates by metal-organic vapour-phase epitaxy (MOVPE) using
standard sources at low pressure in a horizontal reactor setup.
QDs for photoluminescence (PL) measurements were deposited on a
III-V buffer consisting of a strained
In$_{0.06}$Ga$_{0.94}$As/GaAs superlattice between bulk GaAs
layers, as shown in Figure~1a. The strain introduced by the
supperlattice interacts with dislocations and suppresses
propagation into overlying layers. Temperature cycle steps during
buffer growth, performed in a range between 200$^{\circ}$C and
790$^{\circ}$C\@, further support dislocation reduction. QDs for
electroluminescence (EL) experiments were grown on a more
simplified buffer, solely consisting of a GaAs layer of 1 $\mu$m
thickness. On top of each particular buffer, InP QDs were
deposited, embedded in (Al$_{0.5}$Ga$_{0.5}$)$_{0.51}$In$_{0.49}$P
and (Al$_{0.2}$Ga$_{0.8}$)$_{0.51}$In$_{0.49}$P layers serving as
barrier material, lattice matched to GaAs.

For atomic force microscopy (AFM) investigations, QD structures
were grown without capping layers. Figure~1b shows an AFM scan of
a QD layer on top of the extended buffer structure. We mainly find
lens-shaped structures of two different sizes. While the smaller
ones are found all over the investigated area, the bigger ones are
predominantly arranged in chains along the edges of trenches or
valleys. This ordering phenomenon can be explained by effective
local strain fields\cite{Ugur09} at these locations leading to a
preferred nucleation and assembling of InP. We will see later that
the structures observed can be identified as two different types
of QDs. By means of AFM analysis we can estimate the density of
small QDs, in the following named 'type-A' QDs, to be
(4.2$\pm$1)$\times$10$^{10}$\,cm$^{-2}$ and for the large QDs
('type-B') we find (1.9$\pm$1)$\times$10$^9$\,cm$^{-2}$.

The optical properties of the QDs were examined by investigations
of the ensemble luminescence (see Methods). The
temperature-dependent spectra are shown in Figure~2a. At 4.6\,K,
QD emission is observable as broad emission peaks around 1.70\,eV
(type-B) and 1.88\,eV (type-A). At low temperatures, shallow
localisation centres in the barrier material may trap the
optically excited charge carriers. With increasing temperature
these charge carriers can be thermally activated and so both types
of QDs yield a higher intensity. A further increase in temperature
then leads to a decreasing emission intensity, since charge
carriers are thermally activated out of the QD and get lost for
radiative recombination. Due to the stronger quantisation, this
loss process is more severe for the smaller sized type-A QDs,
having a smaller energetic distance between ground state and
barrier. Type-B QDs have a smaller energy-level spacing and
therefore more energy levels in the QD, which may lead to a higher
population number of carriers. In addition, carrier capture and
relaxation processes are more efficient in type-B QDs, enabling
luminescence up to 300\,K\@. The emission characteristics and the
quantum efficiency of the optically pumped QDs, as shown so far,
are in good agreement with the findings for QDs grown on GaAs
substrate\cite{Schulz09}.

The temperature-dependent ensemble luminescence from electrically
pumped InP QDs grown on Si substrate is shown in Figure~2b. Again
we see two types of QDs contributing to the emission at low
temperatures, observable as two broad peaks at 1.73\,eV and
1.91\,eV\@. The slight shift of the peak positions compared to the
optically pumped samples is caused by the doping of the structure,
which causes a bending of the band structures due to the increased
presence of free charge carriers. The emission of type-A QDs can
be observed up to 80\,K, and type-B QDs again show superior
temperature stability even up to room temperature.

In order to confirm the correlation between the structures
observed by AFM and the PL results, luminescence maps of the
optically pumped QD sample were recorded (see Methods), as shown
in Figure~3. The detected light was spectrally filtered, so that
either the spectral range with dominant type-A (590\,nm~-~700\,nm,
Fig.~3b) or type-B (695\,nm~-~785\,nm, Fig.~3c) QD emission could
be investigated. Both maps show exactly the same section of the
sample surface. Comparing these two figures, we clearly see
different luminescence patterns for the different spectral ranges.
The pattern originated by the long-wavelength luminescence centres
(Fig.~3c) fits well to the distribution and dimension of trenches
found by AFM investigations, and around which an agglomeration of
large structures was found (Fig.~1b). The smaller-sized type-A
QDs, which were found evenly spread all over the sample, emit in
the spectral range detected in Figure~3b. In accordance with AFM
results, centres of high luminescence were found evenly
distributed throughout the investigated area.

Furthermore, samples were also subjected to $\mu$-EL measurements
(see Methods). Figure~4a depicts the emission characteristics of a
single QD while varying the bias voltage. Starting at 2\,V, we see
a narrow line at 1.889\,eV, which can be attributed to an
excitonic transition, i.e.\ the recombination of an electron-hole
pair. When more charge carriers are injected, it is getting more
likely for a QD to capture further electrons and holes and form
charged excitons or biexcitons. Increasing the voltage to 2.02\,V,
and thereby also increasing the injection current, an additional
line emerges with an energy difference of 5\,meV\@. This value
fits well to the exciton-biexciton binding energy of approx.\
4-6\,meV in this material system\cite{Beirne07} and thereby
indicates the zero-dimensionality of the light emitter. Figure~4b
shows the temperature-dependent electrically pumped emission from
a single QD. Luminescence can be observed up to 60\,K, keeping up
with electrically pumped InP QDs  grown on GaAs
substrates\cite{Reischle08}. The electrical properties of
structures realised on both types of substrates are as well in
accordance, as apparent in Figure~4c. Here we see the diode
characteristics of InP QD-based LEDs, on Si and GaAs,
respectively. LEDs on Si show a low voltage drop and the same
characteristics like mature LEDs in the GaAs material system.

Since light emitted by single QDs shows significant non-classical
characteristics, second-order autocorrelation measurements
g$^{(2)}$($\tau$) were performed in a Hanbury-Brown and Twiss-type
setup\cite{HBT00} to further verify single dot
emission\cite{Michler00b}. The number of coincidence events versus
the delay time $\tau$ is shown in Figure~5a and b for optically
pumped QDs under pulsed excitation and in Figure~5c for an
electrically pumped QD under DC current excitation. All
measurements show a pronounced suppression of coincidences at
$\tau=0$. The QD in Fig.~5a reveals a g$^{(2)}(0)$ value of 0.08
at 4\,K, which indicates a decrease in multiphoton emission events
by a factor of approximately 12 when compared to a Poissonian
source of the same average intensity. The deviation from
g$^{(2)}$(0)=0, as would be expected for an ideal source, is
caused by the limited temporal resolution of the experimental
setup and by uncorrelated background emission. The latter is
becoming more severe with increasing temperature, as apparent in
the PL spectrum at 80\,K (Fig.~5b). Nevertheless, autocorrelation
measurements showed single-photon emission with g$^{(2)}(0)=0.37$.
Considering the contributions of the background
emission\cite{Reischle08}, we obtained a corrected value for
g$^{(2)}(0)$ of 0.15 at 80\,K\@. Although measurements on
optically pumped QDs were carried out under pulsed excitation, no
correlation peaks can be identified at 4\,K (Fig.~5a). This
behaviour can be explained by charge traps in the barrier
material, in which electrons and holes are stored after optical
excitation. After the decay of the exciton in the QD, the charge
carriers diffuse into the QD and thus populate it again before the
next excitation pulse\cite{Aichele04}. This refilling effect
becomes less pronounced at higher temperatures, as the traps are
then less populated. Consequently, at 80\,K, correlation peaks are
observable at multiples of the laser repetition rate (Fig.~5b).

The autocorrelation measurement on the electrically pumped QD, as
displayed in Figure~5c, shows also non-classical characteristics,
with a g$^{(2)}(0)$ value of 0.52. Here, background correction
yields only a slight reduction down to g$^{(2)}(0)=0.48$, which
indicates that the temporal resolution of the measurement setup is
probably the most constraining factor in this case.

Since we did not deconvolute the instrument response function, all
g$^{(2)}(0)$ values given here should be understood as an upper
limit. With this first presentation of electrically driven
single-photon emission on CMOS-compatible Si substrates, InP QDs
have proven to be highly attractive light sources for future Si
based photonic integrated circuits and quantum information
technology.

\section*{Methods}

For luminescence measurements, the samples were placed in a
He-flow cold finger cryostat. A heater inside the cryostat enabled
temperature control from 4\,K to 300\,K\@. We used a
frequency-doubled Nd:YAG laser for optical excitation in
continuous wave (cw) mode at 532\,nm with a power density of
650\,Wcm$^{-2}$ on the sample. On samples with electrically pumped
QDs, Cr/Zn/Au/Pt/Au-stripes were evaporated to form the p-contact,
and an Al-layer deposited on the back of the sample provided the
n-contact. A current source with a resolution of 0.1\,mA was used
for excitation. For $\mu$-EL measurements, a 50$\times$ microscope
objective collected the EL emission from an area in the order of a
few $\mu$m$^2$. Two stepper motors allowed horizontally and
vertically moving the cryostat, with an effective spatial
resolution of 50\,nm in each direction. The density of optically
active QDs was sufficiently low to investigate single dot emission
without the need of shadow masks.

Recording of the luminescence maps was performed in a liquid
helium bath cryostat. Illumination and detection of the sample was
carried out through a 63$\times$ microscope objective, with the
laser spot being moved by a galvo-scanner. A pinhole was placed in
the detection light path in order to reduce the investigated area
down to a spot size diameter of approx.\ 0.5\,$\mu$m.

\section*{Acknowledgements}
The authors would like to thank E.~Kohler for technical assistance
with the MOVPE, A.~Fuoss for sample preparation, M.~Ubl for device
processing, D.~Richter for AFM measurements, J.~Kettler for
recording luminescence maps and T.~Schwarzb\"ack for illustrative
artwork.


\begin{figure*}[h]
\begin{centering}
\includegraphics[width=0.9\textwidth]{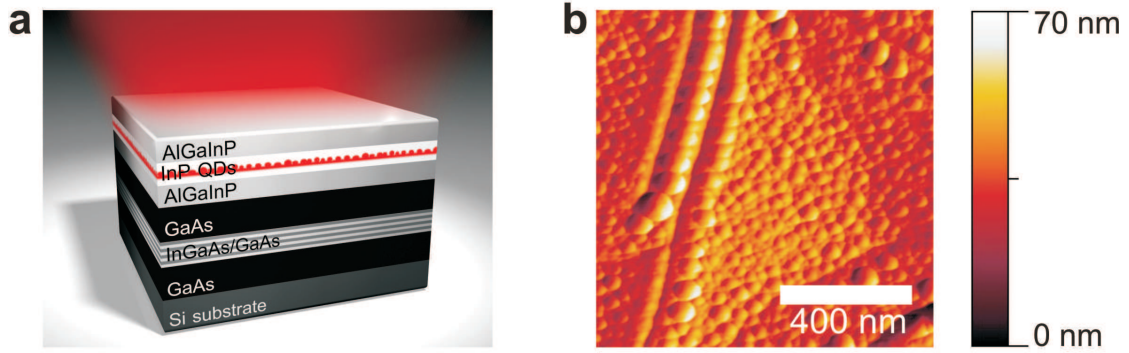}
\caption{\textbf{Sketch of the sample structure and AFM height
measurement.} \textbf{a,} Optically pumped InP quantum dots were
deposited on a buffer consisting of GaAs layers and a
strained-layer superlattice of 10 periods
In$_{0.06}$Ga$_{0.94}$As/GaAs. The superlattice thereby acts as a
dislocation filter. AlGaInP-barriers enclose the QDs and provide
effective charge carrier confinement. \textbf{b,} AFM image of a
QD-layer grown on top of the same structure as depicted in
(\textbf{a}), however without capping layers. The topology of the
investigated section of a scale of 1\,$\mu$m$^2$ exhibits an
asperity with a difference in height of up to 70\,nm.}
\label{Fig1}
\end{centering}
\end{figure*}

\begin{figure*}[h]
\begin{centering}
\includegraphics[width=0.6\textwidth]{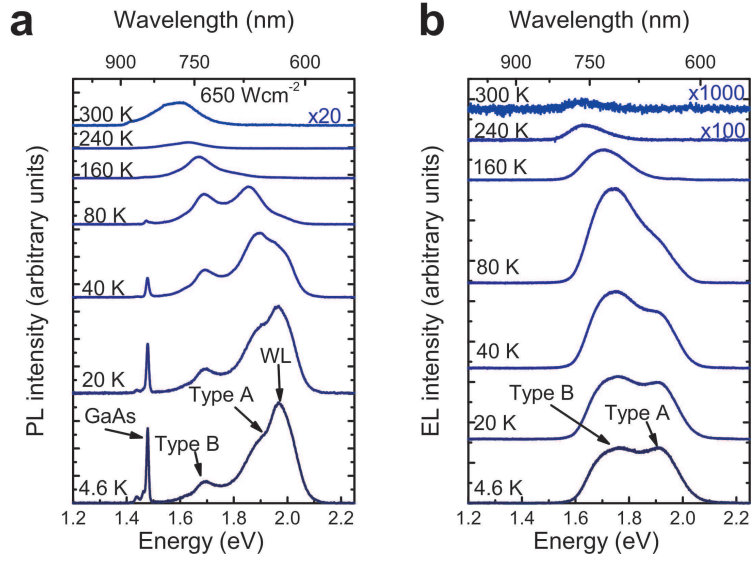}
\caption{\textbf{Temperature dependent ensemble luminescence.}
\textbf{a,} Spectra of optically excited InP-QDs, showing the
emission from the buffer (GaAs), wetting layer (WL) and two types
of QDs (type A and B). \textbf{b,} Spectra of electrically pumped
QDs, taken at a driving current of 50\,mA with the applied voltage
decreasing from 3.35\,V at 4\,K to 1.74\,V at 300\,K\@. Note that
in the case of electrical excitation, neither WL nor GaAs
luminescence is observable, as charge carriers are injected
bidirectional and recombine mainly in the QD layer.} \label{Fig2}
\end{centering}
\end{figure*}

\begin{figure*}[h]
\begin{centering}
\includegraphics[width=0.9\textwidth]{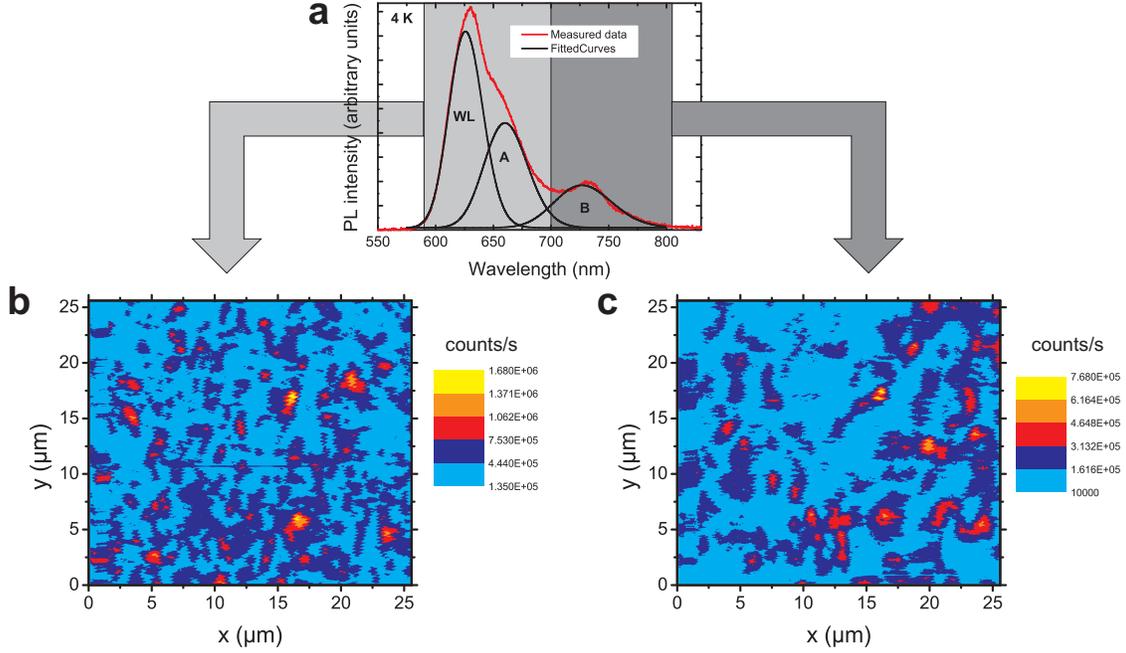}
\caption{\textbf{Spatial PL mapping.} \textbf{a,} Ensemble PL
spectrum of the investigated sample at 4 K. The grey boxes
illustrate the different spectral detection windows. Note that the
emission from type A and B QDs partially overlaps and therefore
can not be completely separated. \textbf{b and c,} PL maps showing
the intensity (colour coded) for two different wavelength regions
of the same surface section. Different colour scales were used for
better visibility. We see different intensity patterns with
apparently less light emitting centres in (\textbf{c}). }
\label{Fig3}
\end{centering}
\end{figure*}

\begin{figure*}[h]
\begin{centering}
\includegraphics[width=0.9\textwidth]{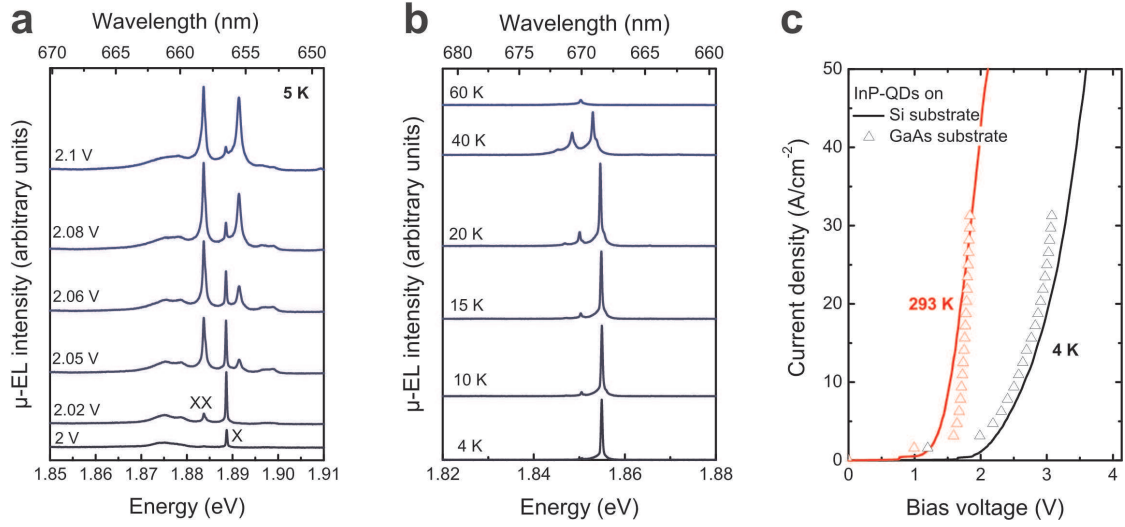}
\caption{\textbf{$\mu$-EL measurements.} \textbf{a,} Emission from
a single QD with increasing pump power. We can identify the
emission of exciton (X) and biexciton (XX). Starting at an applied
voltage of 2.05\,V, an additional line emerges at 1.891\,eV, which
presumably originates from the emission of an adjacent high-energy
QD or a charged exciton state. \textbf{b,} Temperature dependent
EL from a single QD. All spectra were recorded at a bias voltage
of 2.04\,V\@. \textbf{c,} Diode characteristics of QD-based LEDs
grown on Si (solid lines) and on GaAs substrates (symbols). }
\label{Fig4}
\end{centering}
\end{figure*}

\begin{figure*}[h]
\begin{centering}
\includegraphics[width=0.9\textwidth]{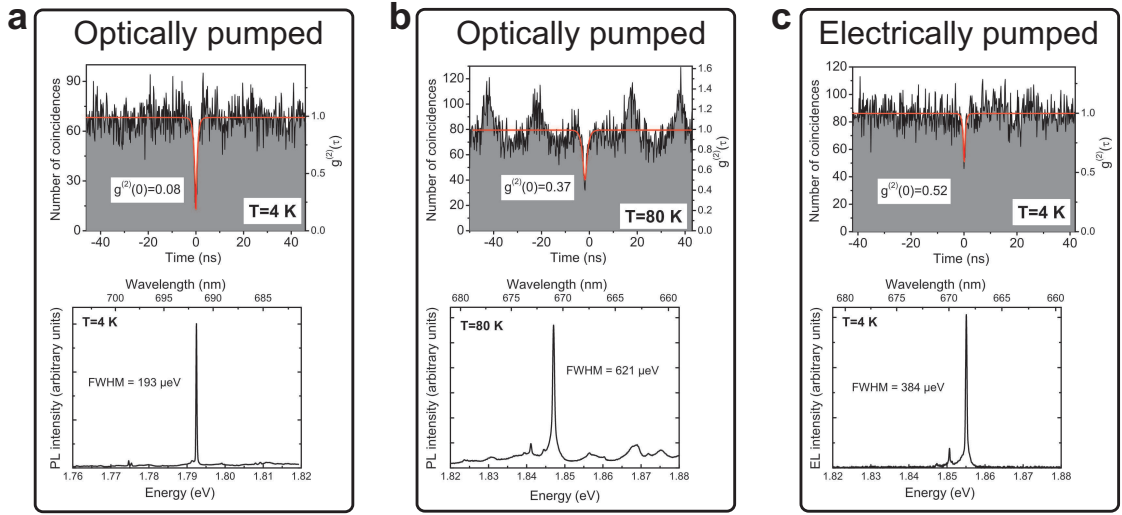}
\caption{\textbf{Second-order autocorrelation measurements.} No
background correction or deconvolution of the instrument response
function was applied. \textbf{a and b,} Measurements on optically
pumped QDs under pulsed excitation at 4\,K and 80\,K, performed on
the excitonic emission lines depicted underneath. Fast refilling
processes, probably caused by charge traps, wash out the
correlation peaks, which would be expected for pulsed excitation.
At 80\,K the charge traps are less populated, hence refilling
effects decline. \textbf{c,} DC current autocorrelation
measurements on the electrically pumped sample show a very narrow
dip due to the limited temporal resolution of the measurement
setup.} \label{Fig5}
\end{centering}
\end{figure*}

\end{document}